\documentclass[preprint,preprintnumbers,amsmath,amssymb,superscriptaddress]{revtex4}
\usepackage{txfonts}
\usepackage{graphicx}
\usepackage{dcolumn}
\usepackage{bm}
\usepackage{array}

\begin{document}

\preprint{APS/123-QED}

\title{Quantum Criticality in Electron-doped BaFe$_{2-x}$Ni$_{x}$As$_{2}$}

\author{R. Zhou}
\author{Z. Li}
\author{J. Yang}
\affiliation{Beijing National Laboratory for Condensed Matter Physics,\\
Institute of Physics, Chinese Academy of Sciences, Beijing 100190, China}
\author{D. L. Sun}
\author{C. T. Lin}
\affiliation{Max Planck Institute-Heisenbergstrasse 1, D-70569 Stuttgart, Germany}
\author{Guo-qing Zheng}
\affiliation{Beijing National Laboratory for Condensed Matter Physics,\\
Institute of Physics, Chinese Academy of Sciences, Beijing 100190, China}
\affiliation{Department of Physics, Okayama University, Okayama 700-8530, Japan}

\maketitle

{\bf
A quantum critical point (QCP) is a point in a system's phase diagram at which an order is completely
suppressed at absolute zero temperature ($T$). The presence of a quantum critical point
manifests itself in the finite-$T$ physical properties, and often gives rise to new states of matter.
Superconductivity in the cuprates and in heavy fermion materials is believed by many
to be mediated by fluctuations associated with a quantum critical point.
In the recently-discovered iron-pnictide high temperature superconductors, 
it is unknown whether a QCP exists or not in a carrier-doped system.
Here we report transport and nuclear magnetic resonance (NMR)   measurements on 
 BaFe$_{2-x}$Ni$_x$As$_2$ (0$\leq x \leq$ 0.17).
We find two critical points at $x_{\rm c1}$ = 0.10 and $x_{\rm c2}$ = 0.14.
 The electrical resistivity follows $\rho ={{\rho }_{0}}+A{{T}^{n}}$,  with $n$ = 1  around $x_{\rm c1}$
and another minimal $n$ = 1.1 at $x_{\rm c2}$. By NMR measurements, we identity $x_{\rm c1}$  to be a  magnetic QCP and suggest that $x_{\rm c2}$ is a new type of QCP associated with a  nematic structural  phase transition.
Our results suggest that the superconductivity in carrier-doped pnictides is   closely linked to the  quantum criticality.}


\vspace{1cm}

In  cuprate high temperature superconductors \cite{LSCO} and heavy fermion \cite{CePd2Si2} compounds, the superconductivity is accompanied by the  normal-state properties deviated from a Landau-Fermi liquid. Such deviation has been ascribed to the quantum critical fluctuations associated with a QCP \cite{Gegenwart_np,Coleman_nat}, whose   relationship to the occurrence of  superconductivity  has been one of the central issues in condensed-matter physics in the last decades.  In  iron-pnictide high temperature superconductors \cite{Hosono,122-1}, searching for magnetic fluctuations  has also become an important subject  \cite{Oka}. 

Quantum critical fluctuations of order parameters take place not only in spatial domain, but also in imaginary time domain \cite{Hertz,Chacravarty}.
The correlation time $\tau_0$ and correlation length $\xi$ are scaled to each other, through a dynamical exponent $z$, $\tau_0\propto \xi^z$.
Several physical quantities, such as the electrical resistivity and spin-lattice relaxation rate (1/$T_1$), can be used to probe  the quantum critical phenomena. For the quasiparticles scattering dominated by the quantum critical fluctuations, the resistivity scales as $\rho\propto T^n$.
For a two-dimensional (2D) antiferromagnetic spin-density-wave (SDW) QCP, the exponent  $n$ = 1 is often observed  \cite{LSCO,CePd2Si2,Kasahara0}. 
On the other hand, for a 2D  order with $q$ = (0,0) where $z$ = 3, $n$ = $\frac{4}{3}$ at the QCP \cite{Moriya_SCR, Xu}. Around a QCP, 1/$T_1$ also shows a characteristic $T$-scaling \cite{Moriya_SCR}.

BaFe$_{2-x}$Ni$_{x}$As$_{2}$ is an electron-doped system \cite{XuZA}
where every Ni donates two electrons in contrast to Co doping that contributes only one electron \cite{Safet}. Therefore, Ni doping suffers less from disorder which is usually harmful for a QCP to exist. In this work, we find two critical points at $x_{\rm c1}$ = 0.10 and $x_{\rm c2}$ = 0.14, respectively. By  NMR measurements, we identify $x_{\rm c1}$ to be a magnetic QCP and suggest that $x_{\rm c2}$  is a QCP associated with the Tetragonal-to-Orthorhombic  structural phase transition. The highest $T_{\rm c}$ is found around  $x_{\rm c1}$, which suggests that  the superconductivity in the carrier-doped BaFe$_2$As$_2$ is more closely tied to the magnetic QCP, while the unusual quantum criticality associated with the structural transition deserves further investigation.

\vspace{1cm}

\textbf{Results}

\begin{figure}
\includegraphics[width=9cm]{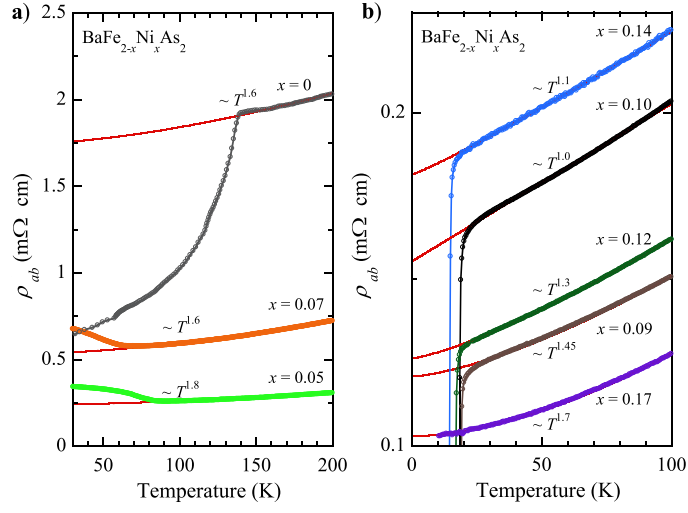}
\caption{\textbf{The in-plane electrical resistivity}. The  in-plane electrical resistivity ${{\rho }_{ab}}$ for BaFe$_{2-x}$Ni$_x$As$_2$. The data in the  normal state are fitted by the equation $\rho_{ab} ={{\rho }_{0}}+A{{T}^{n}}$ (red curve).}
\label{RT_fit}
\end{figure}

\begin{figure}
\includegraphics[width=10cm]{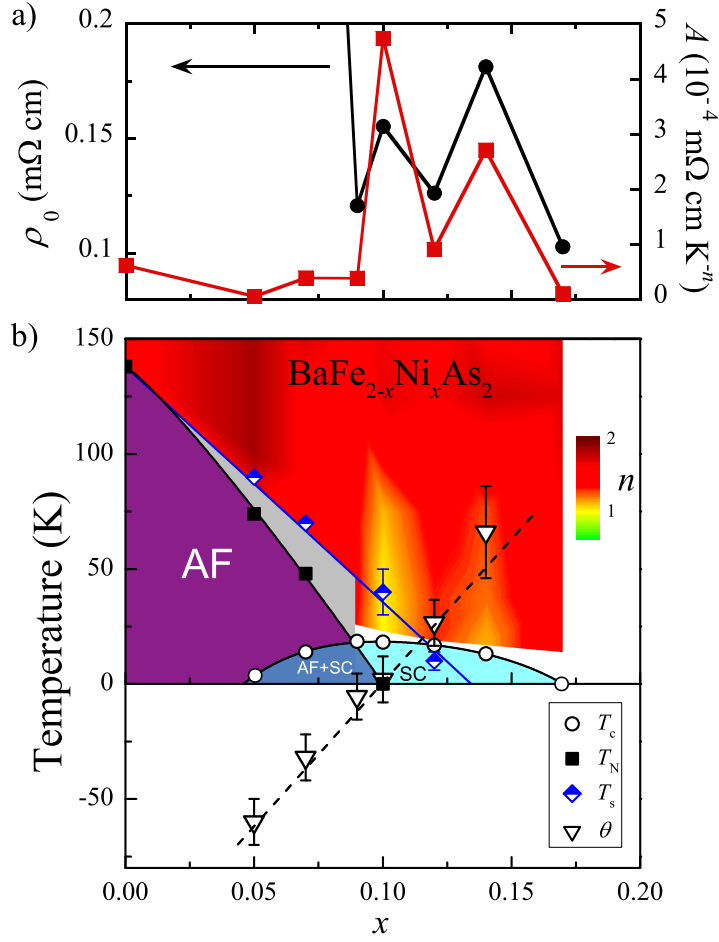}
\caption{ \textbf{The obtained phase diagram of BaFe$_{2-x}$Ni$_x$As$_2$.}  (\textbf{a}) The residual resistivity ${\rho }_{0}$ and the coefficient $A$ as a function of Ni-doping $x$. (\textbf{b}) The obtained phase diagram of BaFe$_{2-x}$Ni$_x$As$_2$. The $T_{\rm N}$ and $\theta$ are obtained from NMR spectra and $1/T_1T$, respectively.
AF and SC denote the antiferromagnetically  ordered and the superconducting states, respectively.
The solid and dashed lines are the guides to the eyes. Colors in the normal state represent the evolution of the exponent $n$ in the resistivity fitted by $\rho_{ab} ={{\rho }_{0}}+A{{T}^{n}}$. The light yellow region at $x$ = 0.10 shows that  the resistivity is $T$-linear.
The error bar for $\theta$ is the standard deviation in the fitting of $1/T_1T$. The error bar for $T_{\rm s}$ represents the temperature interval in measuring the NMR spectra.
}
\label{phase}
\end{figure}

\textbf{Electrical Resistivity measurements.}
Figure \ref{RT_fit} shows the in-plane electrical resistivity data in BaFe$_{2-x}$Ni$_{x}$As$_{2}$  for various $x$, which are fitted by the equation $\rho_{ab} ={{\rho }_{0}}+A{{T}^{n}}$ (for data over the whole temperature range, see Supplementary Figure S1 and Supplementary Note).
For a conventional metal described by the Fermi liquid theory, the exponent $n$ = 2 is expected. However, we find $n$ $<$ 1.5 for $0.09\le x\le 0.14$. This is a notable feature of non-Fermi liquid behavior. Most remarkably, a $T$-linear behavior ($n$ = 1) is  observed for $x_{\rm c1}$ = 0.10 and persists up to $T$ = 100 K.
Another minimal $n$ = 1.1 is found for $x_{\rm c2}$ = 0.14, which is in good agreement with  a previous transport measurement \cite{mu_B_NC}.
An equally interesting feature is that both the residual resistivity $\rho_0$ and the coefficient $A$ show a maximum at $x_{\rm c1}$ and $x_{\rm c2}$ as seen in Fig. \ref{phase} (a).
The evolution of the exponent $n$ with Ni content is shown  in Fig. \ref{phase} (b).

\begin{figure}
\includegraphics[width=10cm]{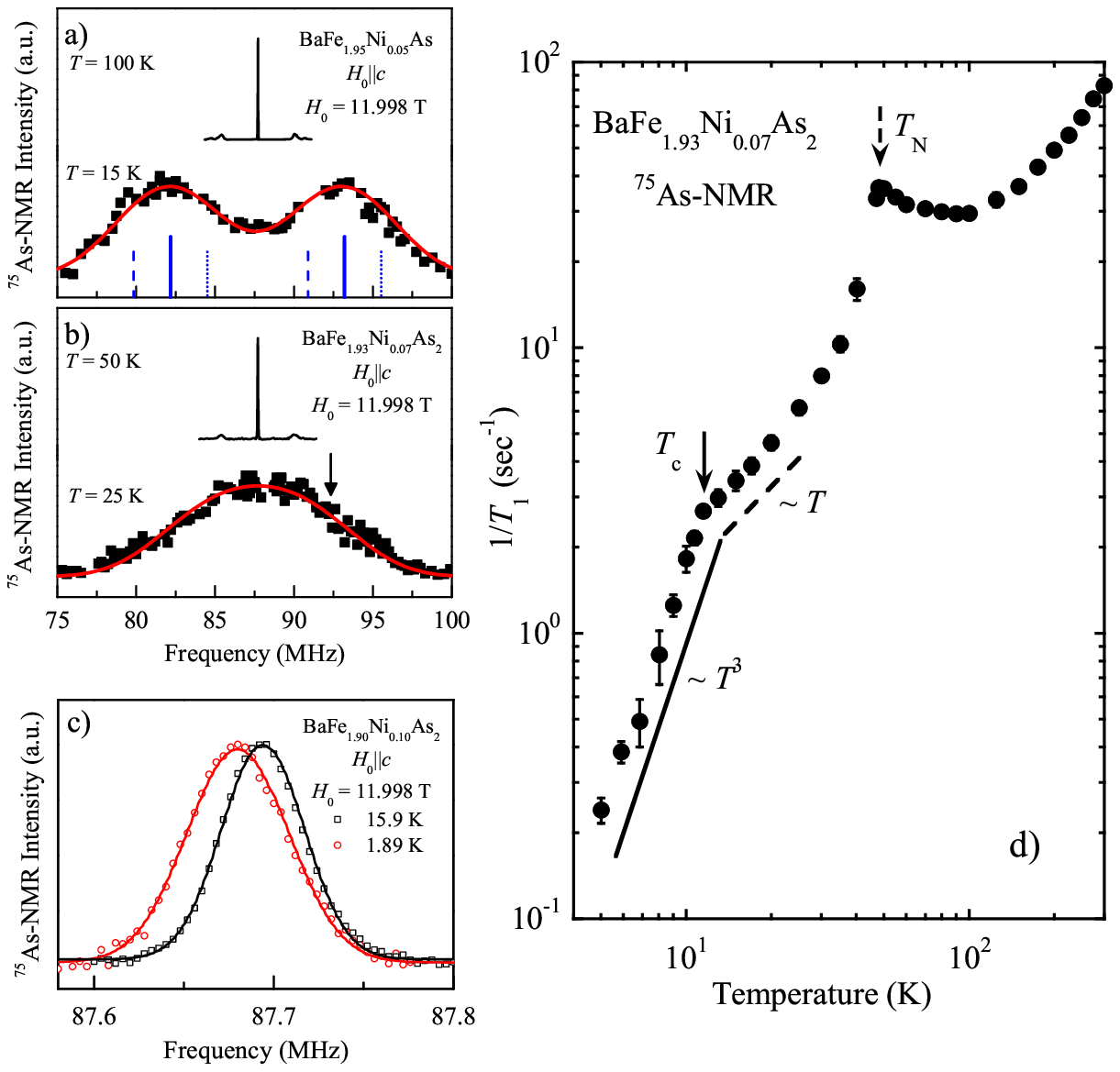}
\caption{\textbf{ The NMR spectra for the underdoped and optimally-doped samples and the 1/$T_1$ for $x$ = 0.07.} $^{75}$As-NMR spectra obtained by sweeping the frequency at a fixed external field $H_0$ = 11.998 T applied along the $c$-axis.  (\textbf{a}) The spectra above and below $T_{\rm N}$ = 74 K for $x$ = 0.05.  Solid, dashed and dotted lines indicate respectively the position of the central transition, left- and right-satellites, when an internal magnetic field develops. (\textbf{b}) The spectra above and below $T_{\rm N}$ = 48 K for $x$ = 0.07.
The red curve represents the simulations by assuming a Gaussian distribution of the internal magnetic field. The black arrow indicates the position at which $T_1$ is measured below $T_{\rm N}$.
(\textbf{c}) The central transition line at and below $T_{\rm c}$ for $x = 0.10$. The spectrum shift at $T$ = 1.89 K is due to a reduction of the Knight shift in the superconducting state. 
(\textbf{d}) The temperature dependence of 1/$T_1$ for $x = 0.07$. The solid and dashed lines show the $T^3$- and $T$-variation, respectively. The solid and dashed arrows indicate $T_{\rm c}$ and $T_{\rm N}$, respectively. The error bar in $1/T_1$ is the standard deviation in  fitting  the 	nuclear magnetization recovery curve.
}
\label{spectra}
\end{figure}

\begin{figure}
\includegraphics[width=9cm]{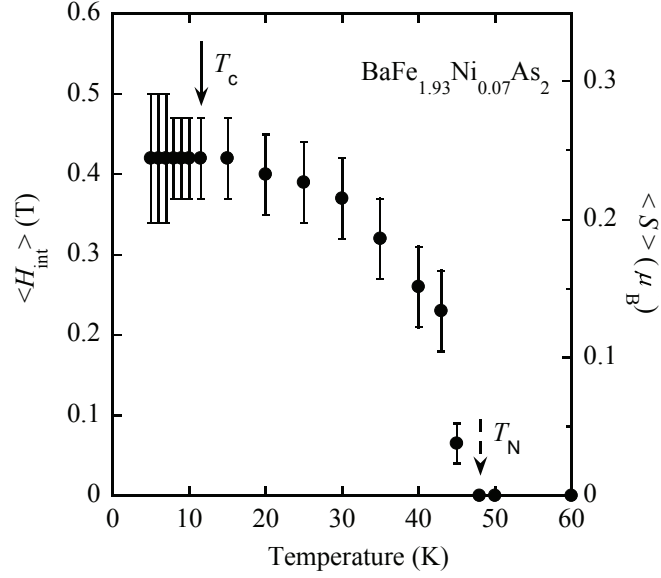}
\caption{\textbf{The temperature dependence of the internal magnetic field and the ordered moment for $x$ = 0.07.} Temperature dependence of the averaged internal magnetic field $\left\langle {{H}_{\rm int}} \right\rangle$ (left vertical axis) and the averaged ordered moment $\left\langle {S} \right\rangle$ (right vertical axis) for x = 0.07. The solid and dashed arrows indicate $T_{\rm c}$ and $T_{\rm N}$, respectively. The error bar represents the absolute  maxima (minima) that can fit  a  spectrum.
}
\label{Hint}
\end{figure}

\textbf{Nuclear magnetic resonance measurements.}
We use NMR to investigate the nature of  $x_{\rm c1}$ and $x_{\rm c2}$. We identify  $x_{\rm c1}$ to be a SDW QCP and suggest that $x_{\rm c2}$ is a Tetragonal-to-Orthorhombic  structural phase transition QCP.
Figure \ref{spectra} (a) and (b) display the frequency-swept spectra of $^{75}$As NMR for $x$ = 0.05 and 0.07, respectively. The very narrow central transition peak above the antiferromagnetic transition temperature $T_{\rm N}$ testifies a good quality of the samples.
When a magnetic order sets in, the spectra will split into two pairs as labeled in Fig. \ref{spectra} (a), due to the development of an internal magnetic field $H_{\rm int}$.
For $x$ = 0.05, 
we observed  two split broad peaks   below $T_{\rm N}$ = 74 K.
  This is due to a distribution of ${H}_{\rm int}$, 
which results in a  broadening of each pair of the peaks.
Upon further doping, ${H}_{\rm int}$ 
is reduced and its distribution  
becomes larger, so that only one broad peak can be seen below $T_{\rm N}$ = 48 K for $x$ = 0.07. 
Similar broadening of the spectra was also observed previously in a very lightly-doped sample BaFe$_{1.934}$Ni$_{0.066}$As$_2$ \cite{Ni-NMR}.
For both $x$, the spectra can be reproduced by assuming a Gaussian distribution of ${H}_{\rm int}$ 
as shown by the red curves. We obtain the averaged internal field $\left\langle {{H}_{\rm int}} \right\rangle $ = 0.75 T at $T$ =15 K for $x$ = 0.05, and $\left\langle {{H}_{\rm int}} \right\rangle $ = 0.39 T at  $T$ = 25 K for $x$ = 0.07.
By using a hyperfine coupling constant of 1.88 T/$\mu_B$  obtained in the undoped parent compound \cite{Kitagawa_BaFe2As2}, the averaged ordered magnetic moment $\left\langle {S} \right\rangle$ is deduced.
As seen in Fig. \ref{Hint}, the  ordered magnetic moment develops continuously below $T_{\rm N}$, being consistent with the second-order nature of the phase transition.
It saturates to 0.43 ${{\mu }_{\rm B}}$ for $x$ = 0.05, and 0.24 ${{\mu }_{\rm B}}$ for $x$ = 0.07, respectively. 
The ordered moment is smaller than that in the hole-doped Ba$_{1-x}$K$_x$Fe$_2$As$_2$ \cite{Li-AF+SC}, which is probably due to the fact  that Ni goes directly into the Fe site and is more effective in suppressing the magnetic order. Upon further doping, at $x$ = 0.09 and $x$ = 0.10, however, no antiferromagnetic transition was found, as demonstrated in Fig.  \ref{spectra} (c) which shows no splitting or broadening ascribable to a magnetic ordering.

The onset of the magnetic order in the underdoped regime is also clearly seen in the spin-lattice relaxation.
Figure \ref{spectra} (d) shows
$1/{{T}_{1}}\ $ for $x$ = 0.07, which  was measured at the position indicated by the arrow in Fig. \ref{spectra} (b),
in order to avoid any influence from possible remnant paramagnetic phase, if any. As seen in Fig. \ref{spectra} (d), a clear peak is found at $T_{\rm N}$ = 48 K due to a critical slowing down of the magnetic  moments. Below $T_{\rm N}$, $1/{{T}_{1}}\ $ decreases down to
$T_{\rm c}$.
Most remarkably, $1/{{T}_{1}}\ $ shows a further rapid decrease below $T_{\rm c}$, exhibiting a $T^3$ behavior down to $T$ = 5 K. Such a significant decrease of $1/{{T}_{1}}\ $ just below $T_{\rm c}$ is due to the superconducting gap opening in the antiferromagnetically ordered state. This is clear and direct evidence for a microscopic coexistence of superconductivity and antiferromagnetism, since the nuclei being measured experience an internal magnetic field yet the relaxation rate is suppressed rapidly below $T_{\rm c}$.
For
$x$ = 0.10, as mentioned already, no antiferromagnetic transition was found. Namely $T_{\rm N}$ = 0.  
This is consistent with the extrapolation of the  $T_{\rm N}$ vs $x$ relation that gives a critical point that coincides with $x_{\rm c1}$ = 0.10, which is further supported by the spin dynamics as elaborated below. 

\begin{figure}
\includegraphics[width=8cm]{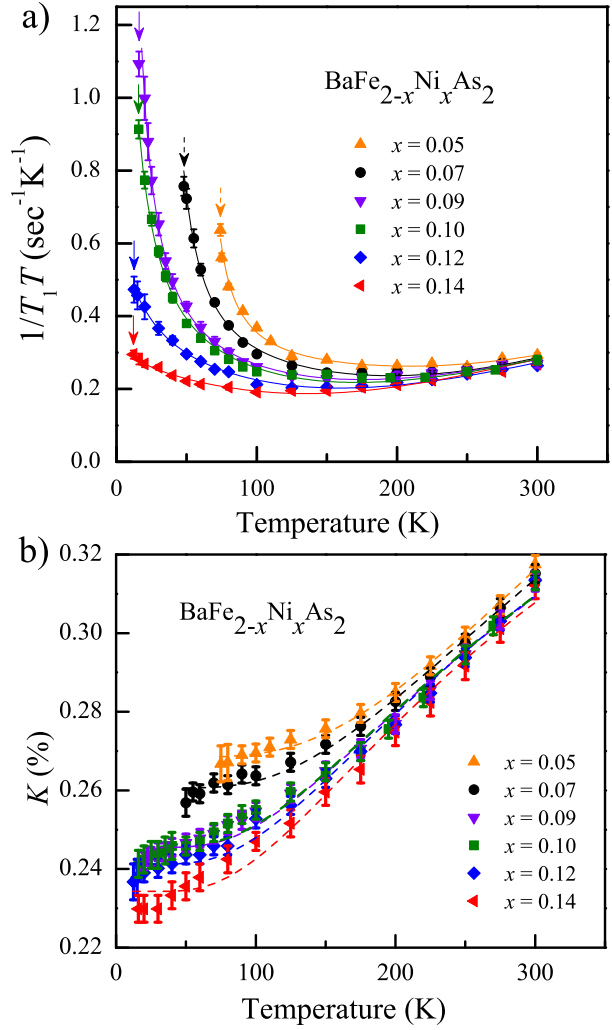}
\caption{ \textbf{The quantity $1/{{T}_{1}}T$ and the Knight shift.} (a) temperature dependence of $1/{{T}_{1}}T$  and (b) the Knight shift $K$   for various $x$.
The solid line is a fitting of $\frac{1}{{{T}_{1}}T}=\frac{a}{T+\theta }+b+c\times \exp \left( -2{{E}_{\text{g}}}/{{k}_{\rm B}}T\  \right)$.
The dashed curve is a fit to $K={{K}_{0}}+K_{\rm s}\times \exp \left( -{{E}_{\rm g}}/{{\rm k}_{\rm B}}T\  \right)$. The obtained parameters are   ${{{E}_{\rm g}}}/{{{k}_{\rm B}}}\;$ = 620$\pm $40, 510$\pm $40, 370$\pm $30, 365$\pm $30, 350$\pm $30 and 330$\pm $30 K for $x$ = 0.05, 0.07, 0.09, 0.10, 0.12 and 0.14, respectively. The solid and dashed arrows indicate $T_{\rm c}$ and $T_{\rm N}$, respectively.
The error bar for $1/{{T}_{1}}T$ is the standard deviation in  fitting  the 	nuclear magnetization recovery curve.
The error bar for $K$ was estimated by assuming that  the spectrum-peak uncertainty equals the point (frequency) interval 	in measuring the NMR spectra.
}
\label{T1T}
\end{figure}
\begin{figure}
\includegraphics[width=9cm]{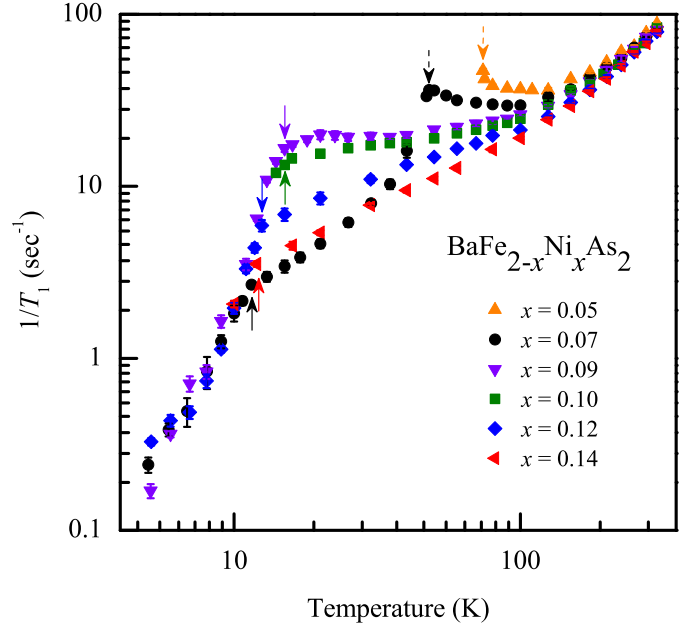}
\caption{ \textbf{The 1/$T_1$ for all the superconducting samples.} The temperature dependence of the spin-lattice relaxation rate 1/$T_1$ for various $x$ of BaFe$_{2-x}$Ni$_x$As$_2$. The solid and dashed arrows indicate $T_{\rm c}$ and $T_{\rm N}$, respectively. The error bar is the standard deviation in  fitting  the nuclear magnetization recovery curve.}
\label{T1}
\end{figure}

 Figure \ref{T1T} (a)  shows the quantity $1/{{T}_{1}}T$  for $0.05\le x\le 0.14$. The $1/{{T}_{1}}T$ decreases with increasing $T$ down to around 150 K, but starts to increase towards $T_{\rm N}$ or $T_{\rm c}$. The increase at low $T$ is due to  the antiferromagnetic spin fluctuation (${{\left( \frac{1}{{{T}_{1}}T} \right)}_{\rm AF}}$) , and the decrease at high $T$ is due to an intraband effect (${{\left( \frac{1}{{{T}_{1}}T} \right)}_{\rm intra}}$).
Namely, $\frac{1}{{{T}_{1}}T}\text{=}{{\left( \frac{1}{{{T}_{1}}T} \right)}_{\rm AF}}\text{+}{{\left( \frac{1}{{{T}_{1}}T} \right)}_{\rm intra}}$.
A similar behavior was also seen in Ba(Fe$_{1-x}$Co$_x$)$_2$As$_2$ \cite{Co-NMR-prl_Ning}. We analyze the ($1/T_1T$)$_{\rm AF}$ part by the self-consistent renormalization (SCR) theory  for a 2D itinerant electron system near a QCP \cite{Moriya_SCR}, which predicts that $1/{{T}_{1}}T$ is proportional to the staggered magnetic susceptibility ${\chi }''\left( Q \right)$.
Since ${\chi }''\left( Q \right)$ follows a Curie-Weiss law \cite{Moriya_SCR}, one has ${{\left( \frac{1}{{{T}_{1}}T} \right)}_{\rm AF}}=\frac{a}{T+\theta }$.
The intraband contribution is due to the density of state (DOS) at the Fermi level, which is related to the spin Knight shift ($K_{\rm s}$) through the Korringa relation $K_{\rm s}^{2}{{T}_{1}}T$ = constant \cite{Korringa}.
The Knight shift was found to follow a $T$-dependence of ${K}=K_0 +  K_{\rm s} \exp \left( -{{E}_{\rm g}}/{{k}_{\rm B}}T \right)$ as seen in Fig. \ref{T1T} (b), where ${{K}_{0}}$ is $T$-independent,
while the second is due to the band that sinks below  the Fermi level \cite{Ikeda,Tabuchi}.
Correspondingly, we can write ${{\left( \frac{1}{{{T}_{1}}T} \right)}_{\rm intra}}=b+c\times \exp \left( -{2{E}_{\rm g}}/{{k}_{\rm B}}T\  \right)$. 
The resulting $\theta$ is plotted in Fig. \ref{phase} (b).
Note that $\theta$ is almost zero at $x$ = 0.10, which yields a constant $1/T_1$ above $T_{\rm c}$ as seen in Fig. \ref{T1}.
The result of $\theta$ = 0 means that the staggered magnetic susceptibility diverges at $T$=0, indicating that $x$ = 0.10 is a magnetic QCP.
Therefore, the exponent $n$ =1 in the resistivity is due to the magnetic QCP
 (Generally speaking,  scatterings due to the magnetic hot spots  give a $T$-linear resistivity \cite{Moriya_SCR}, but those by other parts of the Fermi surface will not \cite{Rice,Fujimoto}.  In  real materials, however, there always exist some extent of impurity scatterings  that can connect the magnetic hot spots and other parts of the Fermi surface as to restore the $n$ =1 behavior at a magnetic QCP).


Next, we use $^{75}$As NMR to study the  structural phase transition. A Tetragonal-to-Orthorhombic structural transition was found in the parent compound \cite{Huang},
but no direct evidence for such structural transition was obtained in the  doped BaFe$_{2-x}$Ni$_{x}$As$_{2}$ thus far.
In BaFe$_{2-x}$Co$_{x}$As$_{2}$, 
a structural transition was detected in the low-doping region \cite{Pratt}, but it is unclear how the transition temperature $T_{\rm s}$ would evolve as doping level increases.
The $^{75}$As nucleus has a nuclear quadrupole moment that couples to the electric field-gradient (EFG) ${{V}_{xx}}$ ($\alpha =x, y, z$). Therefore, the $^{75}$As NMR spectrum  is  sensitive to a  structural phase transition, since  below 
$T_{\rm s}$ the EFG will  change appreciably.
Such change 
was indeed confirmed  in the parent compounds BaFe$_2$As$_2$ \cite{Kitagawa_BaFe2As2} and LaFeAsO \cite{Imai}. When a magnetic field $H_0$ is applied in the $ab$ plane, the NMR resonance frequency $f$ is  expressed by
\begin{equation}
\label{f}
\begin{aligned}
& {{f}_{m\leftrightarrow m-1}}\left( \varphi, \eta  \right) = {{f}_{0}}+\frac{1}{2}{{\nu }_{\rm Q}}\left( m-\frac{1}{2} \right)\times \left(\eta \cdot \cos 2\varphi  -1 \right)
\end{aligned}
\end{equation}
where $m$ = 3/2, 1/2 and -1/2, $\varphi $ is the angle between $H_0$ and the $a$-axis, ${{\nu }_{\rm Q}}$ is the nuclear quadrupole resonance frequency which is proportional to the EFG, and $\eta \equiv \frac{\left| {{V}_{xx}}-{{V}_{yy}} \right|}{{{V}_{zz}}}$. 
For a tetragonal crystal structure,
 $\eta = 0$.
For an orthorhombic structure, however, the  $a$-axis and $b$-axis are not identical, which  results in an asymmetric EFG so that $\eta > 0$.
Therefore, for a twined single crystal,  the  field configurations of
 ${{H}_{0}}\parallel $ $a$-axis ($\varphi $ = ${{0}^{\circ }}$) and ${{H}_{0}}\parallel $ $b$-axis ($\varphi $ = ${{90}^{\circ }}$) will give a different  ${{f}_{m\leftrightarrow m-1}}\left( \varphi, \eta \right)$, 
leading to a splitting of a pair of the satellite peaks into two. The above argument also applies to the case of electronic nematic phase transition such as orbital ordering, since the EFG is also sensitive to a change in the occupation of the on-site electronic orbits.

\begin{figure}
\includegraphics[width=8cm]{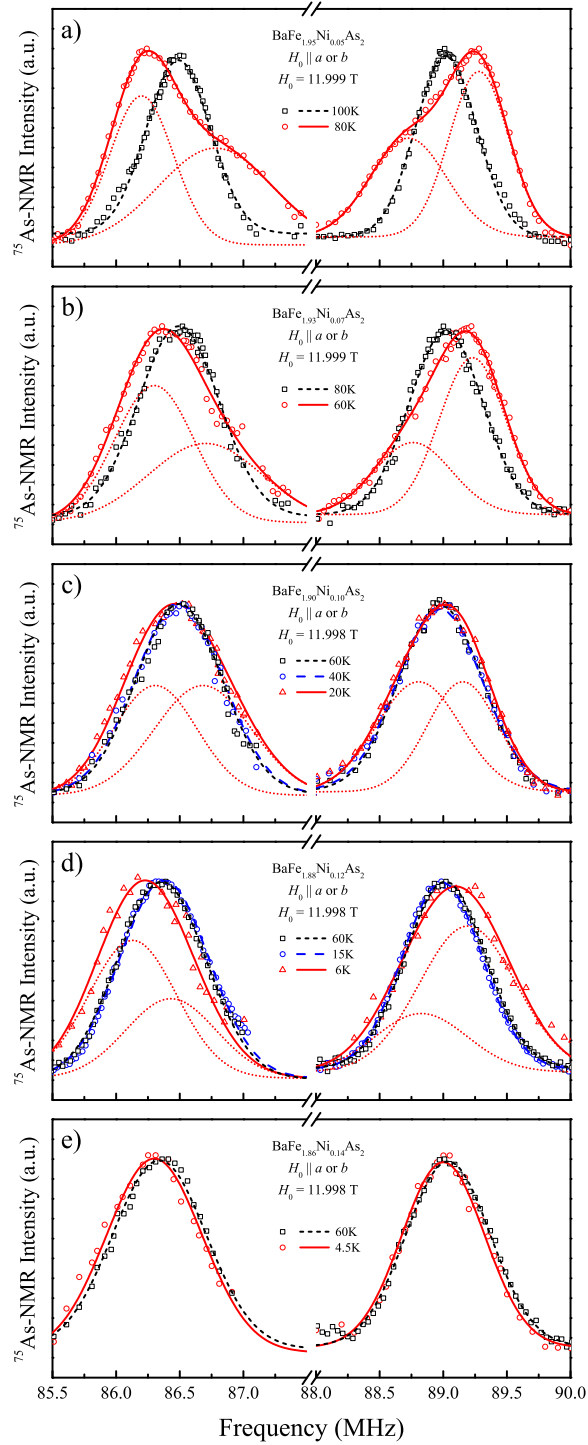}
\caption{ \textbf{The satellite peaks corresponding to the $\frac{1}{2}\leftarrow\rightarrow\frac{3}{2}$ and $-\frac{3}{2}\leftarrow\rightarrow-\frac{1}{2}$ transitions.} (\textbf{a})-(\textbf{d}): the satellite peaks with the external field $H_0$ applied along the $a$-axis or $b$-axis above and below $T_{\rm s}$ for $x$ = 0.05, 0.07, 0.10 and  0.12.  The  peaks  split  below $T_{\rm s}$ due to a change in EFG.
All the spectra  are fitted by a single Gaussian function above $T_{\rm s}$, but   by two Gaussian functions below $T_{\rm s}$.
(\textbf{e}): the satellite peaks for $x$ = 0.14 which only shift to the same direction due to a reduction of the Knight shift as in Fig. \ref{spectra} (\textbf{c}).}
\label{spectra_ts}
\end{figure}

\begin{figure}
\includegraphics[width=10cm]{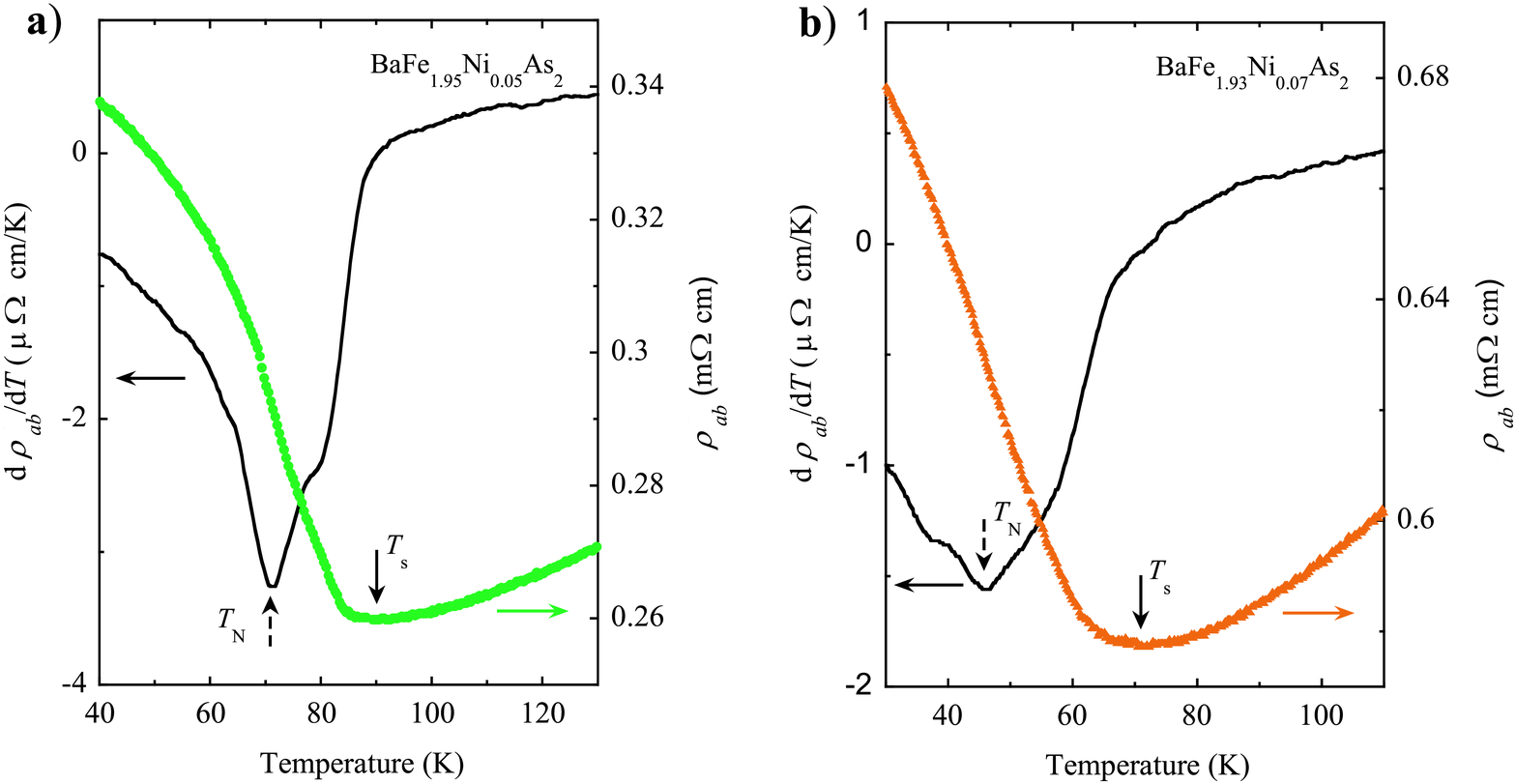}
\caption{ \textbf{The enlarged part  of the electrical resistivity and its derivative for the underdoped samples.} \textbf{(a)} Enlarged part of the in-plane electrical resistivity $\rho_{ab}$ (right vertical axis) and its derivative (left vertical axis) for $x$ = 0.05. \textbf{(b)} Data for $x$=0.07. The solid and dashed arrows indicate the structural phase transition temperature $T_s$ and the antiferromagnetic transition temperature $T_N$ determined by NMR, respectively.}
\label{d_rt}
\end{figure}

As shown in Fig. \ref{spectra_ts} (a) and (b),  only one pair of
satellite peaks is observed at high temperature for both $x$ = 0.05 and 0.07.  Below certain temperature, $T_{\rm s}$, however, we observed  a splitting of the satellite peaks.
This is strong microscopic evidence   for a structural transition occurring in the underdoped samples,  where the NMR spectra split owing to the formation of the twinned orthorhombic domains. In fact, each satellite peak can be well reproduced by assuming two split peaks.
The obtained $T_{\rm s}$ = 90 K for $x$ = 0.05 and $T_{\rm s}$ = 70 K for $x$ = 0.07 agrees well with the resistivity data where an upturn at $T_{\rm s}$ is observed,  see Fig. \ref{d_rt}.
This feature in the spectra persists to the higher dopings $x$ = 0.10 and 0.12.
For the optimal doping $x$ = 0.10, the spectrum is  nearly unchanged for  40 K $\leq T \leq$ 60 K, but   suddenly changes at 20 K. This indicates that a structural transition takes place below $T_{\rm s}\sim$ 40 K for this doping composition.
For $x$ = 0.12, the two satellite peaks do not change at high temperatures, but shift to the opposite direction below $T$ = 10 K. Each broadened peak at $T$ = 6 K can be well fitted by a superposition of two peaks. This is the first observation that a structural phase transition takes place below $T_{\rm c}$. For $x$ = 0.14, however, no
broadening of the spectra was found down to $T$ = 4.5 K; the spectrum shift to the same direction is due to a reduction of the Knight shift in the superconducting state. This indicates that $T_{\rm s}$ = 0 around $x_{\rm c2}$ = 0.14.
The $T_{\rm s}$ results obtained by NMR are summarized in the phase diagram as shown in Fig. \ref{phase} (b).

   Therefore, the minimal $n$=1.1 of the resistivity exponent around $x_{\rm c2}$ implies quantum critical fluctuations associated with the structural QCP.
We hope that our work will stimulate more experimental measurements such as elastic constant which is also sensitive to such critical fluctuations \cite{Yoshizawa}.
It should be emphasized here that the  exponent $n$ = 1.1 is smaller than $n$ = $\frac{4}{3}$  expected for the fluctuations dominated by $q$ = (0,0) \cite{Xu}. Our result suggests that  the fluctuation associated with $x_{\rm c2}$ is local (namely, all wave vectors contribute to the quasiparticles scattering), which would lead to an $n$ = 1 \cite{Miyake}.


\vspace{1cm}

\textbf{Discussion}

The mechanism for the superconductivity in  iron-pnictides has been discussed in relation to magnetic fluctuations \cite{Mazin,Kuroki,Scalapino} as well as structural/orbital fluctuations \cite{Kontani}. In this context,
it is  worthwhile noting that  $T_{\rm c}$ is the highest around  $x_{\rm c1}$. 
This  suggests that the superconductivity in BaFe$_{2-x}$Ni$_x$As$_2$ is more closely related to the antiferromagnetic QCP rather than the structural QCP. Also, previous results suggestive of a magnetic QCP was  reported for BaFe$_2$(As$_{1-x}$P$_x$)$_2$\cite{P_PRL_Ishida,Kasahara}, but isovalent P substitution for As does not add carriers there. In the present case, Ni directly donates electrons into the system (chemical doping), so the tuning parameter is totally different here. Thus, our work demonstrates that BaFe$_{2-x}$Ni$_{x}$As$_2$ is a new material that provides a unique opportunity to study the issues of quantum criticality. 
In particular, our result suggests that $x_{\rm c2}$ is a new type of QCP at which the exponent $n$ cannot be  explained by   existing theories. Although the importance of   the Tetragonal-to-Orthorhombic structural transition, which is often associated with electronic nematicity \cite{Chu,Shen}, has been pointed out in the pnictides \cite{Fernandes,Kontani,Ono}, the physics of a QCP associated with it is a much less-explored frontier, which deserves more investigations in the future.



\vspace{1cm}

\textbf{Methods}

\textbf{Sample preparation and characterization:} The single crystal samples of BaFe$_{2-x}$Ni$_{x}$As$_2$ used for  the  measurements were grown by the self-flux method \cite{crystal grow}. Here, the Ni content $x$ was determined by energy-dispersive x-ray spectroscopy (EDX). The $T_{\rm c}$ was determined by DC susceptibility measured by a superconducting quantum interference device with the applied field 50 Oe parallel to the $ab$ plane. The $T_{\rm c}$ is 3.5, 14, 18.5, 18.2, 16.8 and 13.1 K for $x$ = 0.05, 0.07, 0.09, 0.10, 0.12 and 0.14, respectively.

\textbf{The resistivity measurements:} Resistivity measurements were performed in Quantum Design physical properties measurement system (PPMS) by a standard dc four-probe method. Here we have used the same single crystals used in the NMR measurements. The electrical resistance is measured upon both warming and cooling processes in order to ensure no temperature effect from the electrodes on the samples. Both the warming and the cooling speed is 2 K/min.

\textbf{Measurements of NMR spectra and $T_1$:} The NMR spectra were obtained by integrating the spin echo as a function of the RF frequency at a constant external magnetic field $H_0$ = 11.998 T. The nucleus $^{75}$As has a nuclear spin $I$ = 3/2 and the nuclear spin Hamiltonian can be expressed as
\begin{equation}
\label{H}
\begin{aligned}
& H=\gamma hI\left( {{H}_{0}}+{H}_{\operatorname{int}} \right) \\ &
 +\frac{h{{\nu }_{\rm Q}}}{6}\left[ 3I_{\rm Z}^{2}-I\left( I+1 \right)+\frac{1}{2}\eta \left( I_{+}^{2}+I_{-}^{2} \right) \right]
\end{aligned}
\end{equation}
where ${\gamma }/{2\pi }\;=7.9219\  {\rm MHz}/{\rm T}\;$ is the gyromagnetic ratio of $^{75}$As, $h$ is Planck constant, ${H}_{\rm int}$ is the internal magnetic field at the As nuclear spin site resulting from the hyperfine coupling to the neighboring Fe electrons.
The $T_1$ was determined by using the  saturation-recovery method, and the nuclear magnetization  is fitted to $1-\frac{M\left( t \right)}{M\left( \infty  \right)}=0.9{{\exp }^{-6t/{{T}_{1}}\;}}+0.1{{\exp }^{-t/{{T}_{1}}\;}}$, where $M(t)$ is the nuclear magnetization at time $t$ after the saturation pulse \cite{method_T1}. The curve is fitted very well and $T_1$ is of single component.

\textbf{Simulation of the spectra:}
In order to reproduce the spectra, we assume that the distribution of the internal magnetic field is Gaussian as,
\begin{equation}
\label{I}
\begin{aligned}
I\left( f \right)=\exp \left( -\frac{{{\left\{ f-\left[ \gamma \left( {{H}_{0}}\pm \left\langle {{H}_{\operatorname{int}}} \right\rangle  \right)+n{{\nu }_{c}} \right] \right\}}^{2}}}{2{{H}_{\sigma }}^{2}} \right)
\end{aligned}
\end{equation}

where $I\left( f \right)$ is the intensity of the spectra, $H_{\sigma }$ is the distribution of the internal magnetic field $H_{\rm int}$, and ${{\nu }_{c}}$ is the nuclear quadrupole resonance (NQR) frequency tensor along the $c$-axis. Then the spectrum is fitted by convoluting Eq. \ref{I} with a Gaussian broadening function $p\left( {{f}'} \right)={{A}_{n}}\frac{1}{\sqrt{2\pi }{{\sigma }_{n}}}\exp \left( -\frac{{{{{f}'}}^{2}}}{2{{\sigma }_{n}}^{2}} \right)$ that already exists above $T_{\rm N}$. For $n$ = 0 (central transition), $A_0$ = 1 and $2\sqrt{2\ln 2}{{\sigma }_{0}}$ is the full width at half maximum (FWHM) of the central peak above $T_{\rm N}$. For $n$ =  $\pm 1$ (satellites), ${{A}_{\pm 1}}=\frac{3}{4}$ and $2\sqrt{2\ln 2}{{\sigma }_{\pm 1}}$ is the FWHM of satellite lines above $T_{\rm N}$.
The red solid curve in Fig. \ref{spectra} (a) is a fit with ${{H}_{\sigma }}$ = 0.37 T, $2\sqrt{2\ln 2}{{\sigma }_{0}}$ = 0.03 MHz, and $2\sqrt{2\ln 2}{{\sigma }_{\pm 1}}$ = 0.38 MHz which was taken from the spectrum at $T$ = 100 K. The red solid curve in Fig. \ref{spectra} (b) is a fit with $2\sqrt{2\ln 2}{{\sigma }_{0}}$ = 0.04 MHz and $2\sqrt{2\ln 2}{{\sigma }_{\pm 1}}$ = 0.5 MHz which was taken from the spectrum at $T$ = 50 K.

\textbf{Acknowledgments} We thank Q. Si, K. Miyake, K. Kuroki, T. Takimoto, Y. Yanase, R. Fernandes, T. Xiang and S. Fujimoto for helpful discussion. This work was partially supported by  National Basic Research Program of China (973 Program), Nos. 2011CBA00100 $\And$ 2011CBA00109,  and by CAS.

{\bf Author contributions}
The single crystals were grown by D.L.S and C.T.L. The NMR measurements were performed by R.Z., Z.L., J.Y and  G.Q.Z. The electrical resistivity was measured by  R.Z., Z.L., and J.Y. G.Q.Z coordinated the
whole work and wrote the manuscript, which was supplemented by R.Z. All authors have discussed the results and the interpretation.

{\bf Additional information}
Supplementary information is available in the online version of the paper.
Correspondence and requests for materials should be addressed to G.Q.Z.

{\bf Competing financial interests}
The authors declare no competing financial interests.

\clearpage

\clearpage

\Large{\textbf{Supplementary Figure}}

\vspace{2cm}

\centerline{\includegraphics[width=10cm]{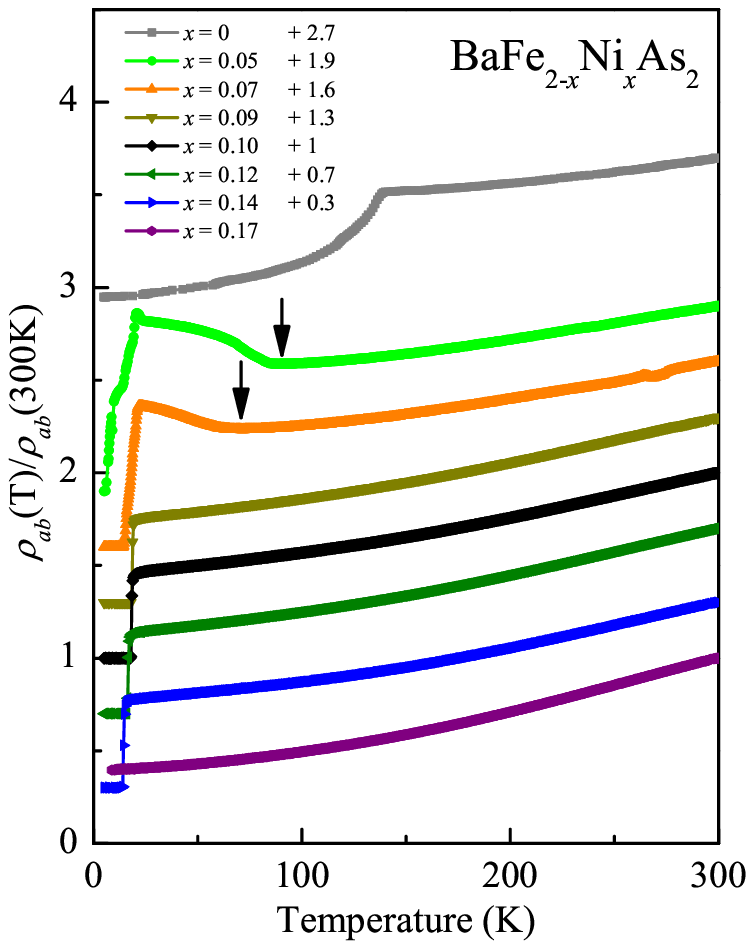}}

\begin{flushleft}
\small \textbf{Supplementary Figure S1: The electrical resistivity in the whole temperature region.} The in-plane electrical resistivity $\rho_{ab}$/$\rho_{ab}$(300 K) for the whole temperature region of BaFe$_{2-x}$Ni$_x$As$_2$ at zero field. Except for the sample $x$ = 0.17, data were offset vertically by a number shown in the figure. The arrows indicate the structural phase transition temperature $T_{\rm s}$ of $x$ = 0.05 and 0.07.
\end{flushleft}

\clearpage

\Large{\textbf{Supplementary Note: Electrical resistivity}}
\vspace{0.5cm}

\normalsize
Supplementary Figure S1 shows the in-plane electrical resistivity $\rho_{ab}$ up to $T$ = 300 K. In our studies, we find that the $T$-linear dependence of the resistivity can only be observed at $x$ = 0.10. Therefore, even for a nominal sample with $x$ = 0.10, if a sample is inhomogeneous, the electrical resistivity will be easily influenced by other compositions and will not show a $T$-linear behavior. This is probably the reason why the $T$-linear behavior of electrical resistivity has never been observed before for the BaFe$_{2-x}$Ni$_x$As$_2$ system\cite{mu_B_NC}.

The residual resistivity is larger than that in Ba$_{1-x}$K$_x$Fe$_2$As$_2$ or BaFe$_2$(As$_{1-x}$P$_x$)$_2$. For example, $\rho_{0}$  = 15.5 $\mu\Omega$ cm in Ba$_{0.68}$K$_{0.32}$Fe$_2$As$_2$[42], and  $\rho_{0}$  = 33 $\mu\Omega$ cm in BaFe$_2$(As$_{0.67}$P$_{0.33}$)$_2$\cite{Kasahara0}. In comparison, $\rho_{0}$ in the Ni-doped samples is larger by more than three times. This is a result that Ni goes directly into the Fe site, which has a stronger disorder effect compared to Ba-site or As-site substitutions. However, since the $A\cdot {{T}^{n}}$ term is quite large, it has little effect on the fitting to obtain the exponent $n$.

\vspace{1cm}

\Large{\textbf{Supplementary References}}
\vspace{0.5cm}

\small
\leftline{[42] ~~Gasparova, V. A., Wolff-Fabrisb, F., Sun, D. L., Lin, C. T.  $\And$ Wosnitzab, J.. Electron Transport and}
  ~~~~Anisotropy of the Upper Critical Magnetic Field in Ba$_{0.68}$K$_{0.32}$Fe$_2$As$_2$ Single Crystals. \emph{JETP Letters}

  ~~~~\textbf{93}, 26-30 (2011).


\begin{references}

%
\bibitem{LSCO}Cooper, R. A.  \emph{et al.} 
    Anomalous criticality in the electrical resistivity of La$_{2-x}$Sr$_x$CuO$_4$. \emph{Science} \textbf{323}, 603-607 (2009).
    %
\bibitem{CePd2Si2}Mathur, N. D.  \emph{et al.} 
    Magnetically mediated superconductivity in heavy fermion compounds.  \emph{Nature} \textbf{394}, 39-43 (1998).

%
\bibitem{Gegenwart_np}Gegenwart, P.,  Si, Q. $\And$ Steglich, F.. Quantum criticality in heavy-fermion metals. \emph{Nature Phys.} \textbf{4}, 186-197 (2008).
%
\bibitem{Coleman_nat}Coleman, P. $\And$ Schofield, A. J.. Quantum criticality. \emph{Nature} \textbf{433}, 226-229 (2005).

\bibitem{Hosono}Kamihara, Y., Watanabe, T., Hirano, M.  $\And$ Hosono, H..  Iron-based layered superconductor La[O$_{1-x}$F$_x$]FeAs ($x$ = 0.05-0.12) with $T_{\rm c}$ = 26 K. \emph{J. Am. Chem. Soc.} \textbf{103}, 3296-3297 (2008).
%
\bibitem{122-1} Rotter, M., Tegel, M.  $\And$ Johrendt, D.. Superconductivity at 38 K in the iron arsenide (Ba$_{1-x}$K$_x$)Fe$_2$As$_2$. \emph{Phys. Rev. Lett.} \textbf{101}, 107006 (2008).

\bibitem{Oka}
Oka, T. \emph{et al.}  
Antiferromagnetic Spin Fluctuations above the Dome-Shaped and Full-Gap Superconducting States of LaFeAsO$_{1-x}$F$_{x}$ Revealed by $^{75}$As-Nuclear Quadrupole Resonance.
\emph{Phys. Rev. Lett.} {\bf 108}, 047001 (2012).

\bibitem{Hertz}
Hertz, J. A.. Quantum critical phenomena. \emph{Phys. Rev. B} \textbf{14}, 1165-1184 (1976).
%
\bibitem{Chacravarty}
Chakravarty, S., Halperin, B. I. $\And$ Nelson, D. R..
Two-dimensional quantum Heisenberg antiferromagnet at low temperatures.
\emph{Phys. Rev. B} \textbf{39}, 2344-2371 (1989).

\bibitem{Kasahara0}	Kasahara, S. \emph{et al.} Evolution from non-Fermi- to Fermi-liquid transport via isovalent doping in BaFe$_2$(As$_{1-x}$P$_x$)$_2$ superconductors. \emph{Phys. Rev. B} {\bf 81}, 184519 (2010).
\bibitem{Moriya_SCR}Moriya, T.. Theory of itinerant electron magnetism. \emph{J. Mag. Mag. Mat.} \textbf{100}, 261-271 (1991).
\bibitem{Xu}
Xu, C.,  Muller, M. $\And$ Sachdev, S..  Ising and spin orders in the iron-based superconductors. \emph{Phys. Rev. B} {\bf 78}, 020501 (2008).
\bibitem{XuZA}
 Li, L. J. \emph{et al.} Superconductivity induced by Ni doping
in BaFe$_2$As$_2$ single crystals. \emph{New J. Phys.} \textbf{11}, 025008 (2009).

\bibitem{Safet}
Sefat, A. S. \emph{et al.} 
Superconductivity at 22 K in Co-Doped BaFe$_2$As$_2$ Crystals.
\emph{Phys. Rev. Lett.} \textbf{101}, 117004 (2008).

\bibitem{mu_B_NC}Luo, H. Q.  \emph{et al.} 
    Coexistence and competition of the short-range incommensurate antiferromagnetic order with the superconducting state of BaFe$_{2-x}$Ni$_x$As$_2$. \emph{Phys. Rev. Lett.} \textbf{108}, 247002 (2012).


\bibitem{Ni-NMR}
Dioguardi, A. P. \emph{et al.}
    Local magnetic inhomogeneities in Ba(Fe$_{1-x}$Ni$_x$)$_2$As$_2$ as seen via $^{75}$As NMR. \emph{Phys. Rev. B} \textbf{82}, 140411 (2010).
%
\bibitem{Kitagawa_BaFe2As2}Kitagawa, K., Katayama, N., Ohgushi, K., Yoshida, M. $\And$ Takigawa, M.. Commensurate itinerant antiferromagnetism in BaFe$_2$As$_2$: $^{75}$As-NMR studies on a self-flux grown single crystal. \emph{J. Phys. Soc. Jpn.} \textbf{77}, 114709 (2008).
%
\bibitem{Li-AF+SC}Li, Z. \emph{et al.} 
 Microscopic coexistence of antiferromagnetic order and superconductivity in Ba$_{0.77}$K$_{0.23}$Fe$_2$As$_2$.
\emph{Phys. Rev. B} {\bf 86}, 180501 (2012).


%
\bibitem{Co-NMR-prl_Ning}Ning, F. L. \emph{et al.} 
    Contrasting spin dynamics between underdoped and overdoped Ba(Fe$_{1-x}$Co$_x$)$_2$As$_2$. \emph{Phys. Rev. Lett.} \textbf{104}, 037001 (2010).
%
\bibitem{Korringa}Korringa, J.. Nuclear magnetic relaxation and resonnance line shift in metals. \emph{Physica} \textbf{16} (7-8), 601-610 (1950).
%
\bibitem{Ikeda}
Ikeda, H.. Pseudogap and superconductivity in iron-based layered superconductor studied by fluctuation-exchange approximation.  \emph{J. Phys. Soc. Jpn.}  {\bf 77},  123707  (2008) .
\bibitem{Tabuchi}Tabuchi, T. \emph{et al.} Evidence for a full energy gap in the nickel pnictide supercon-ductor LaNiAsO$_{1-x}$F$_x$ from $^{75}$As nuclear quadrupole resonance. \emph{Phys. Rev. B} \textbf{81}, 140509 (2010).



\bibitem{Rice}
Hlubina, R. $\And$ Rice, T. M.. Resistivity as a function of temperature for models with hot spots on the Fermi surface. \emph{Phys. Rev. B} {\bf 51}, 9253-9261 (1995).

\bibitem{Fujimoto}
Fujimoto, S., Kohno, H. $\And$ Yamada, K.. Temperature dependence of electrical resistivity in two-dimension Fermi systems. \emph{J. Phys. Soc. Jpn.} {\bf 60}, 2724-2728 (1991).
%
\bibitem{Huang}
Huang, Q. \emph{et al.} Neutron-Diffraction Measurements of Magnetic Order and a Structural Transition in the Parent BaFe$_2$As$_2$ Compound of FeAs-Based High-Temperature Superconductors. \emph{Phys. Rev. Lett.} \textbf{101}, 257003 (2008).
\bibitem{Pratt}Pratt, D. K. \emph{et al.} Coexistence of competing antiferromagnetic and superconducting phases in the underdoped Ba(Fe$_{0.953}$Co$_{0.047}$)$_2$As$_2$ compound using X-ray and neutron scattering techniques. \emph{Phys. Rev. Lett.} \textbf{103}, 087001 (2009).



\bibitem{Imai}
Fu, M. {\it et al.} NMR Search for the Spin Nematic State in LaFeAsO Single Crystal. \emph{Phys. Rev. Lett.} \textbf{109}, 247001  (2012).
 \bibitem{Yoshizawa}
 Yoshizawa, H. {\it et al.},  Structural Quantum Criticality and Superconductivity
in Iron-Based Superconductor Ba(Fe$_{1-x}$Co$_x$)$_2$As$_2$.
\emph{J. Phys. Soc. Jpn.} \textbf{81}, 024604 (2012).

 \bibitem{Miyake}
Watanabe, S. $\And$ Miyake, K.. Quantum valence criticality as an origin of unconventional critical phenomena. \emph{Phys. Rev. Lett.} {\bf 105}, 186403 (2010).

 \bibitem{Mazin}
Mazin, I. I., Singh, D. J., Johannes, M. D. $\And$ Du, M. H.. Unconventional
sign-reversing superconductivity in LaFeAsO$_{1-x}$F$_x$. \emph{Phys. Rev. Lett.} {\bf 101}, 057003 (2008).

\bibitem{Kuroki}
Kuroki, K. {\it et al.} Unconventional pairing originating from the disconnected
Fermi surfaces of superconducting LaFeAsO$_{1-x}$F$_x$. \emph{Phys. Rev. Lett.} {\bf 101}, 087004 (2008).

\bibitem{Scalapino}
Graser, S.  {\it et al.} 
 Spin fluctuations and superconductivity in a three-dimensional tight-binding model for BaFe$_2$As$_2$.
\emph{Phys. Rev. B} {\bf  81}, 214503 (2010).

\bibitem{Kontani}
Kontani, H. $\And$  Onari, S..
 Orbital-Fluctuation-Mediated Superconductivity in Iron Pnictides: Analysis of the Five-Orbital Hubbard-Holstein Model.
\emph{Phys. Rev. Lett.} {\bf 104}, 157001 (2010).
 %
\bibitem{P_PRL_Ishida}Nakai, Y.  \emph{et al.} 
    Unconventional superconductivity and antiferromagnetic quantum critical behavior in the isovalent-doped BaFe$_2$(As$_{1-x}$P$_x$)$_2$. \emph{Phys. Rev. Lett.} \textbf{105}, 107003 (2010).
\bibitem{Kasahara}
Hashimoto, K.  {\it et al.} A Sharp Peak of the Zero-Temperature Penetration Depth at Optimal Composition in BaFe$_2$(As$_{1-x}$P$_x$)$_2$. \emph{Science} {\bf 336}, 1554-1557 (2012).






\bibitem{Chu}
Chu, J.-H. {\it et al.} 
In-Plane Resistivity Anisotropy in an Underdoped Iron Arsenide Superconductor.
\emph{Science} \textbf{329}, 824-826 (2010).

\bibitem{Shen}
Yi, M. {\it et al.} 
Symmetry breaking orbital anisotropy on detwinned Ba(Fe$_{1-x}$Co$_x$)$_2$As$_2$ above the spin density wave transition.
\emph{Proc. Nat. Acad. Sci.}  \textbf{108}, 6878-6883 (2011).

\bibitem{Fernandes}
Fernandes, R. M.,  Chubukov, A. V., Knolle, J.,   Eremin, I.  $\And$ Schmalian, J..
Preemptive nematic order, pseudogap, and orbital order in the iron pnictides.
\emph{Phys. Rev. B} \textbf{85}, 024534 (2012).



\bibitem{Ono}
Yanagi, Y., Yamakawa, Y., Adachi, N. $\And$ Ono, Y..
 Cooperative effects of Coulomb and electron-phonon interactions in the two-dimensional 16-band $d$-$p$ model for iron-based superconductors.
\emph{Phys. Rev. B} \textbf{82}, 064518 (2010).

\bibitem{crystal grow}Sun, D. L., Liu, Y.,  Park, J. T.  $\And$ Lin, C. T.. Growth of large single crystals of BaFe$_{1.87}$Co$_{0.13}$As$_2$ using a nucleation pole. \emph{Supercond. Sci. Technol.} \textbf{22}, 105006 (2009).


\bibitem{method_T1}Narath, A.. Nuclear Spin-Lattice Relaxation in Hexagonal Transition Metals: Titanium. \emph{Phys. Rev.} \textbf{162}, 320-332 (1967).



\end{references}
\end{document}